\documentclass[twocolumn,english,aps,prb,twocolum,superscriptaddress,natbib,bibnotes,amsmath,amssymb,floatfix,groupedaddress,footinbib]{revtex4-2}
\usepackage[colorlinks=true,citecolor=blue,linkcolor=magenta]{hyperref}

\usepackage[markup=nocolor, authormarkupposition=left]{changes} 
\usepackage{soul}
\usepackage[utf8]{inputenc}
\usepackage[english]{babel}
\usepackage{amsmath,amsfonts,amssymb}
\usepackage[T1]{fontenc}
\usepackage{url}
\usepackage{amsmath}
\usepackage{siunitx}
\usepackage[version=4]{mhchem}
\usepackage{bm}

\usepackage{amsfonts}
\usepackage{amssymb}

\usepackage{epstopdf}
\usepackage{graphicx}
\graphicspath{{./Figures/}}

\newcommand{\Fref}[1]{Figure~\ref{#1}}
\newcommand{\fref}[1]{Fig.~\ref{#1}}

\begin{document}

%\title{Low-temperature and \hl{ultra low loss...} hydrogen-free silicon dioxide cladding\\ for next-generation integrated photonics}

%\title{Low-temperature deposition of low-loss silica cladding for next-generation integrated photonics}

\title{Hydrogen-free low-temperature silica for next generation integrated photonics}

\author{Zheru Qiu$^{1,2,\dagger}$, Zihan Li$^{1,2,\dagger}$, Rui Ning Wang$^{1,2,3}$, %
Xinru Ji$^{1,2}$, Marta Divall$^{1,2}$, Anat Siddharth$^{1,2}$, Tobias J. Kippenberg$^{1,2,\ast}$}

\address{${}^1$Swiss Federal Institute of Technology Lausanne (EPFL), CH-1015 Lausanne, Switzerland\\
${}^2$Center for Quantum Science and Engineering, EPFL, CH-1015 Lausanne, Switzerland\\
${}^3$Present Address: Luxtelligence SA, CH-1015, Lausanne, Switzerland}

\maketitle
\pretolerance=2500
\noindent\textbf{%
	The advances in novel low-loss "on insulator" integrated photonics platforms beyond silicon, such as thin-film \ce{LiNbO3}, \ce{LiTaO3}, \ce{GaP} and \ce{BaTiO3} have demonstrated major potential across a wide range of applications, due to their unique electro-optical or nonlinear optical properties.
	This has heralded novel devices, ranging from low-voltage and high-speed modulators to parametric amplifiers.
	For such photonic integrated circuits, a low-loss \ce{SiO2} cladding layer is a key element, serving as a passivation layer for the waveguides and enabling efficient fiber-to-chip coupling.
	However, numerous novel ferroelectric or III-V "on insulator" platforms have low tolerances for process temperature. 
	%, fundamentally limited by the thermal expansion mismatch and material stability.
	This prohibits using high-temperature anneals to remove hydrogen, a common impurity that is inherent to ordinary chemical vapor deposited \ce{SiO2} and causes significant optical loss in the near-infrared.
	Here, we satisfy the dichotomy of a low-loss wafer scale manufactured silicon dioxide cladding and low processing temperature.
	Inspired by the manufacturing of optical fibers, we introduce a hydrogen-free, low-loss \ce{SiO2} cladding that is deposited at low temperatures ($\SI{300}{\degreeCelsius}$) by using \ce{SiCl4} and \ce{O2} as precursors in inductively coupled plasma-enhanced chemical vapor deposition (ICPCVD).
	By replacing hydrogenous silicon precursors (e.g. silane, i.e. SiH4) with silicon tetrachloride \ce{SiCl4}, the deposited film is inherently free from residual hydrogen.
	The low process temperature is compatible with the "on insulator" platforms and CMOS electronic integrated circuits.
	We demonstrate a wide low-loss window that covers all telecommunication bands from $\SI{1260}{\nano\meter}$ to $\SI{1625}{\nano\meter}$.
	We achieve a $\bm<\SI{2.5}{dB/m}$ waveguide loss at $\SI{1550}{\nano\meter}$, comparable with $\SI{1200}{\degreeCelsius}$ annealed films. % state-of-the-art
    % future
	Our \ce{SiCl4} process provides a key future cladding for all recently emerged "on-insulator" photonics platforms, that is low cost, scalable in manufacturing, and directly foundry compatible.
 }
\section{Introduction}
\begin{figure*}[ht]
  \centering
  \includegraphics[width=0.95\textwidth]{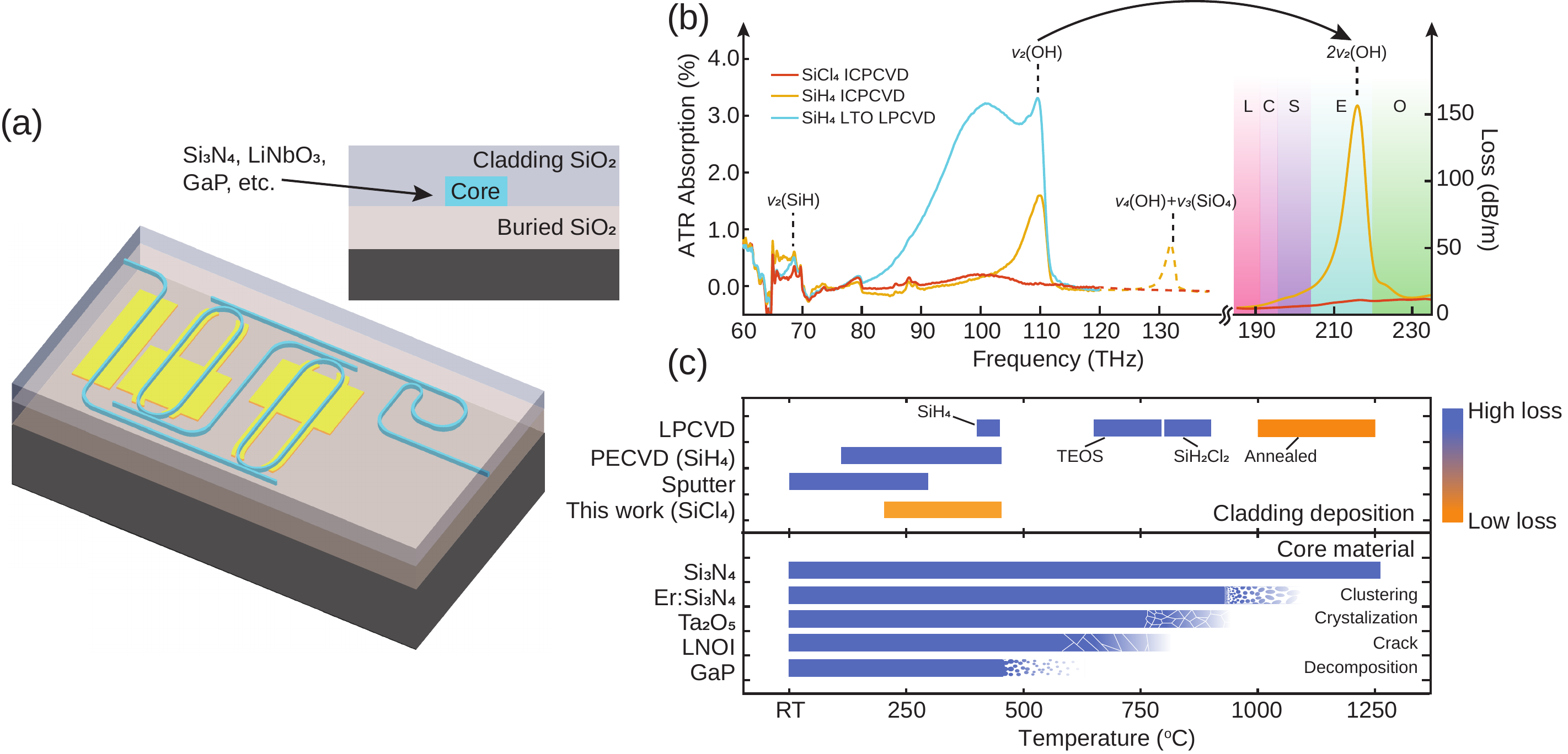}
  \caption{\footnotesize \textbf{Common deposition methods and loss of \ce{SiO2} cladding for integrated photonic circuits.}
    (a) Illustration of an advanced electro-optical active photonic integrated circuit chip with \ce{SiO2} cladding. 
    (b) Near and mid infrared optical absorption of \ce{SiO2} deposited with different processes. Highlighting the telecom band loss originating from the overtone of the \ce{OH} impurity vibration modes.
    (c) Temperature requirement and typical near-infrared loss of the different deposition techniques of \ce{SiO2} film and the temperature tolerance of material platforms.
}
\label{Fig1}
\end{figure*}

Emerging low-loss integrated photonics platforms based on the thin-film "on insulator" crystalline materials such as lithium niobate (\ce{LiNbO3}) \cite{LNOIRev}, lithium tantalate (\ce{LiTaO3}) \cite{LTOI} and barium titanate (\ce{BaTiO3}) \cite{btocleo} are offering unprecedented functionalities, enabling key applications such as agile electro-optic tuning\cite{snigirev2023ultrafast} and efficient nonlinear wavelength conversion\cite{LNOIHybridComb}.
Materials such as \ce{GaP}\cite{GaPSoliton}, \ce{AlGaAs}\cite{AlGaAs} and \ce{Ta2O5} on insulator\cite{Tatanla, TatanlaUCSB} are also attractive because of their strong Kerr nonlinearity.
However, these platforms have a low tolerance for high temperature, fundamentally limited to less than \SI{800}{\degreeCelsius}.
Low-temperature deposition of low-near-infrared loss \ce{SiO2} films for the essential waveguide passivation is a well-known challenge due to the hydrogen impurity from deposition precursors (\fref{Fig1}(b)).
Hours-long $>\SI{1000}{\degreeCelsius}$ annealing is usually required for a low-loss film.
Here we present a novel \SI{300}{\degreeCelsius} plasma-enhanced chemical vapor deposition process for low-loss hydrogen-free \ce{SiO2} using \ce{SiCl4} precursor.
We eliminated the \ce{OH} absorption and reduced the loss in the entire \SI{1260}{\nano\meter} to \SI{1620}{\nano\meter} band, enabling the full capabilities of today’s low-loss integrated photonics platforms.

The utilization of new materials introduced new capabilities but also necessitated low-temperature fabrication processes (\fref{Fig1}(c)).
The common thin-film \ce{LiNbO3} and \ce{LiTaO3} on insulator wafer consists of an ion-sliced thin film and a silicon substrate, the thermal expansion coefficient mismatch between the film and the substrate limits the maximum temperature to ca. \SI{700}{\degreeCelsius} before film delamination or cracking\cite{LNOIRevBowers}.
Many advanced ferroelectric materials have low Curie temperatures --- $\sim$\SI{685}{\degreeCelsius} for \ce{LiTaO3}\cite{LTCurie} --- furthermore limits the tolerable heating.
Many III-V materials are also not tolerant of high temperatures due to the decomposition \cite{IIIVheat3, IIIVheat4} when beyond the growth temperature.
Recrystallization\cite{zhu2006intense} of the amorphous \ce{Ta2O5} above \SI{650}{\degreeCelsius} will also add to the waveguide scattering loss.
Similarly, erbium-doped silicon nitride waveguides are promising for efficient, high-power, high-density on-chip amplification\cite{Erbium}.
The highly doped \ce{Si3N4} active medium allows excellent gain and output power but is susceptible to erbium clustering and performance degradation after extensive annealing\cite{ErClusteringAlt}, also requiring a low-temperature low-loss cladding.

On the other hand, modern low-loss photonic integrated circuits require a cladding as the passivation layer to protect the waveguides and to allow further integration.
Silicon dioxide (\ce{SiO2}) is the ubiquitously used cladding material for its low refractive index and suitable chemical properties.
In the commonly used \ce{SiO2} chemical vapor deposition processes, the gaseous silicon precursors like \ce{SiH4}, \ce{SiH2Cl2}, and \ce{Si(OC2H5)4} (TEOS) all contain hydrogen, and thus lead to hydrogen impurity trapped in the deposited film.
This impurity causes significant optical loss in near-infrared mainly due to the vibration overtone of the \ce{OH} bonds it forms (\fref{Fig1}(b)), which manifests as a strong peak around \SI{1380}{\nano\meter} with a long tail into longer wavelengths\cite{OHBand, humbach1996analysis}.

This notorious absorption can only be removed by anneals at very high temperatures (usually >\SI{1000}{\degreeCelsius})\cite{yongheng2006study}, which will destroy \ce{LiNbO3} waveguides\cite{LNreflow}.
%In the fabrication process of \ce{Si3N4} photonics\cite{SiN}, the cladding layers are typically annealed at \SI{1200}{\degreeCelsius} for \SI{11}{h}.
%This temperature is only $\sim\SI{50}{\degreeCelsius}$ lower than the melting point of \ce{LiNbO3} and can destroy waveguiding structures\cite{LNreflow}.
Furthermore, the removal of hydrogen is a diffusion-controlled process, implying that the required time exponentially increases with the film thickness and can be tens of hours for few-\SI{}{\micro\meter} of \ce{SiO2}.

In literature, replacing the natural hydrogen with deuterium is shown to reduce the \ce{OH} absorption at \SI{1380}{\nano\meter}\cite{D2Fiber, D2Cladding}.
However, deuterium is expensive due to its low abundance (0.016\%) and the energy-intensive isotope separation process, which it requires.
The residual \ce{OH} absorption due to the incomplete isotope separation and the shifted \ce{OD} absorption can also hamper wide-band applications such as frequency comb generation\cite{LNOIEOComb}, parametric amplification and optical computation\cite{tensorcore}.

The pioneers in the silica optical fiber manufacturing also suffered from \ce{OH} absorption, which is solved by the invention of fiber preform preparation processes like the modified chemical vapor deposition (MCVD)\cite{MCVD} and the plasma chemical vapor deposition (PCVD)\cite{PCVD} process.
In these processes, \ce{SiCl4}, as the precursor for \ce{Si}, is oxidized in a strongly heated tube.
As the entire process does not involve any hydrogen, the root cause of the hydrogen contamination is eliminated.
% \hl{ZQ: I noted the comment on SiCl4 is liquid and can be removed of impurities, and reacts with metals to form volatile componds. But I don't think this point applies here as actually it's more for the dopants like Ge in fibers.
% See: https://ieeexplore.ieee.org/abstract/document/1131071
% The process takes advantage of the fact that the vaport pressure of these dopants is many orders of magnitude higher
% than any transition metal impurities which might be present in the starting materials and cause absorption losses thus, the entrained dopants are of very high purity relative to such contaminants. Hydrogenated species, however, typically have very similar vapor pressures as the dopants and are therefore commonly purified to remove such contaminants. }

Here we present an inductively coupled plasma-enhanced chemical vapor deposition (ICPCVD) process inspired by the MCVD and PCVD.
In this process, low-loss \ce{SiO2} free from hydrogen absorption is directly deposited at low temperature ($<$\SI{300}{\degreeCelsius}) with \ce{SiCl4} as the silicon precursor and \ce{O2} as oxidizer (\fref{Fig2}(a)).
Neither of the precursors contains isotopes of hydrogen, which provides an avenue to produce a completely hydrogen-free \ce{SiO2} film at a low cost.
The elimination of the \ce{OH} absorption not only reduces the loss in the technologically important S and C telecommunication bands but also enables operation in the very wide low-loss window spanning the entire \SI{1260}{\nano\meter} to \SI{1620}{\nano\meter} spectrum.
The deposition is performed at temperatures as low as \SI{300}{\degreeCelsius}, which is compatible with CMOS devices\cite{CMOSThermal} and various integrated photonics platforms including \ce{LiNbO3} on insulator.
We believe this advance can become an important part of the manufacture of modern low-loss photonic integrated circuits with stringent requirements on process temperature inherent to material properties.

% latex hack to protect from upper casing
\newcommand{\SiCl}{\ce{SiCl4}}
\newcommand{\SiO}{\ce{SiO2}}
\section{{\protect\SiCl} based {\protect\SiO} deposition process}

\ce{SiCl4} as a silicon compound with high vapor pressure, has been widely used as the \ce{Si} precursor in the fabrication of silica fiber preforms.
It has been utilized in producing thousands of kilometers of telecom-grade low-loss optical fiber, connecting the world by supporting the Internet from data centers to households.
However, replacing the hydrogenous precursors by \ce{SiCl4} in integrated photonic applications is not straightforward.
Although the molecular structure of \ce{SiCl4} molecules is similar to the commonly used silicon precursor \ce{SiH4} in low-temperature depositions, oxidation of \ce{SiCl4} into \ce{SiO2} is much more unfavorable thermodynamically.
The reaction enthalpy $\Delta H_r$ of \ce{SiCl4} oxidation, which is the amount of heat generated under constant pressure per unit amount of \ce{SiO2} deposited, is only 17.5\% of \ce{SiH4} at room temperature\cite{CRC} (see Supplementary). %$\SI{1136}{kJ}$

\begin{figure*}[h]
  \centering
  \includegraphics[width=0.99\textwidth]{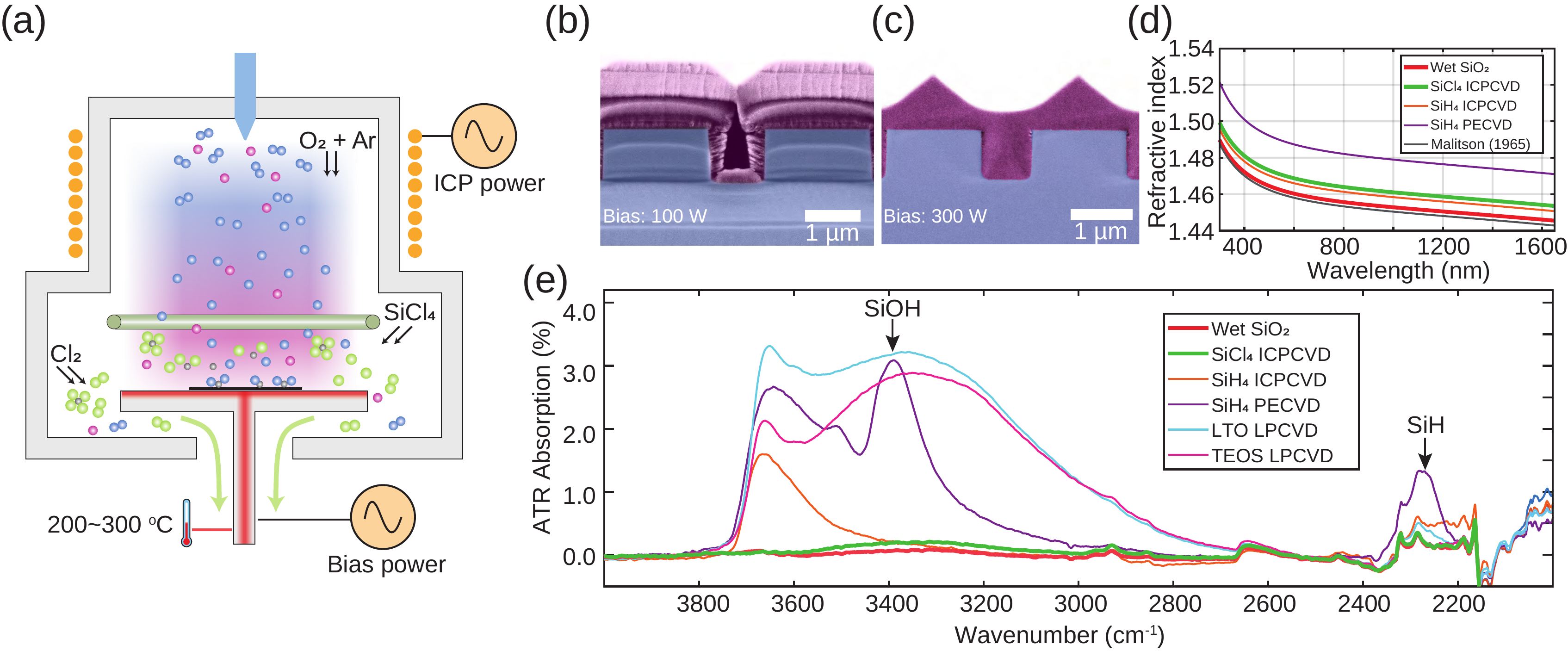}
  \caption{\footnotesize \textbf{Low temperature deposition of high-quality \ce{SiO2} films by \ce{SiCl4} based ICPCVD.}
  (a) Illustration of the ICPCVD reactor for the deposition process.
  (b), (c) False colored cross-section SEM of the deposited \ce{SiO2} on \SI{1}{\micro\meter} \ce{Si} deep trenches with around 1:1 aspect ratio, deposited with bias power of 150~W and 430~W, respectively.%
  (d) Refractive index of different \ce{SiO2} films characterized by ellipsometry, in comparison to the data from \cite{MalitsonSiO2}.%
  (e) Fourier transform infrared (FTIR) absorption spectrum of \ce{SiO2} film deposited by \ce{SiCl4} ICPCVD, in comparison to films created by other methods.
  }
\label{Fig2}
\end{figure*}

In the MCVD and PCVD process for fiber preforms, the mixture of \ce{SiCl4} vapor and \ce{O2} is heated to more than \SI{1150}{\degreeCelsius} to allow gas phase oxidation.
This high reaction temperature defeats the purpose of using hydrogen-free precursors.
To overcome the requirement of high temperature, an inductively coupled radio frequency plasma (ICP) reactor is used in this work, where the high-density plasma makes the reaction possible at much lower temperatures.
The higher plasma density and electron temperature in the ICP reactor\cite{CVDBook} can promote the disassociation of \ce{SiCl4} molecules and accelerate the formation of \ce{SiO2} film. 

\begin{widetext}
	\begin{equation*}
		\begin{aligned}
			\ce{SiH4 (g) + O2 (g) &-> SiO2 (s) + 2H2O (g)},\qquad&\Delta H_r^\circ =\SI{-1377}{kJ/mol}\\
			\\
			\ce{SiCl4 (g) + O2 (g) &-> SiO2 (s) + 2Cl2 (g)},\qquad&\Delta H_r^\circ =\SI{-241}{kJ/mol}
		\end{aligned}
	\end{equation*}
\end{widetext}

For this reason, the new process is demonstrated in a commercial inductively coupled plasma chemical vapor deposition (ICPCVD) reactor (Oxford Instrument PlasmaPro 100, \fref{Fig2}(a)), with a custom \ce{SiCl4} gas line installed.
The typical condition of deposition is around 30~sccm \ce{SiCl4} flow, 40 to 80~sccm \ce{O2} flow, 0 to 30~sccm \ce{Ar} flow, 4 to 8~mTorr total pressure, \SI{300}{\degreeCelsius} table temperature, around 2300~W ICP source power and 100 to 400~W of capacitively coupled radio frequency (RF) power (bias power).
A very high deposition rate of $>\SI{40}{\nm\per\min}$ can be achieved under these conditions.
As the produced film only has a moderate compressive stress of $\sim-250\SI{}{\mega\pascal}$, $>\SI{10}{\micro\meter}$-thick layers can be deposited without cracking.
Beyond the challenge of producing a thin film, for cladding of waveguides, void-free filling of high-aspect-ratio structures is required\cite{shiran2020impact}.
The ion bombardment leads to simultaneous sputtering of the deposited film, slowing down the deposition but improving the step coverage of deposited film and the capability of filling high-aspect-ratio structures (Fig.~\ref{Fig2} (b),(c))\cite{CVDBook}.
The rate of sputtering can be controlled by adjusting the bias RF power capacitively applied to the reactor.
For the cladding of waveguides, a two-step deposition can be applied, one step with high bias power for stronger ion bombardment and another with lower power for a higher deposition rate.

\begin{figure*}[!htb]
  \centering
  \includegraphics[width=0.99\textwidth]{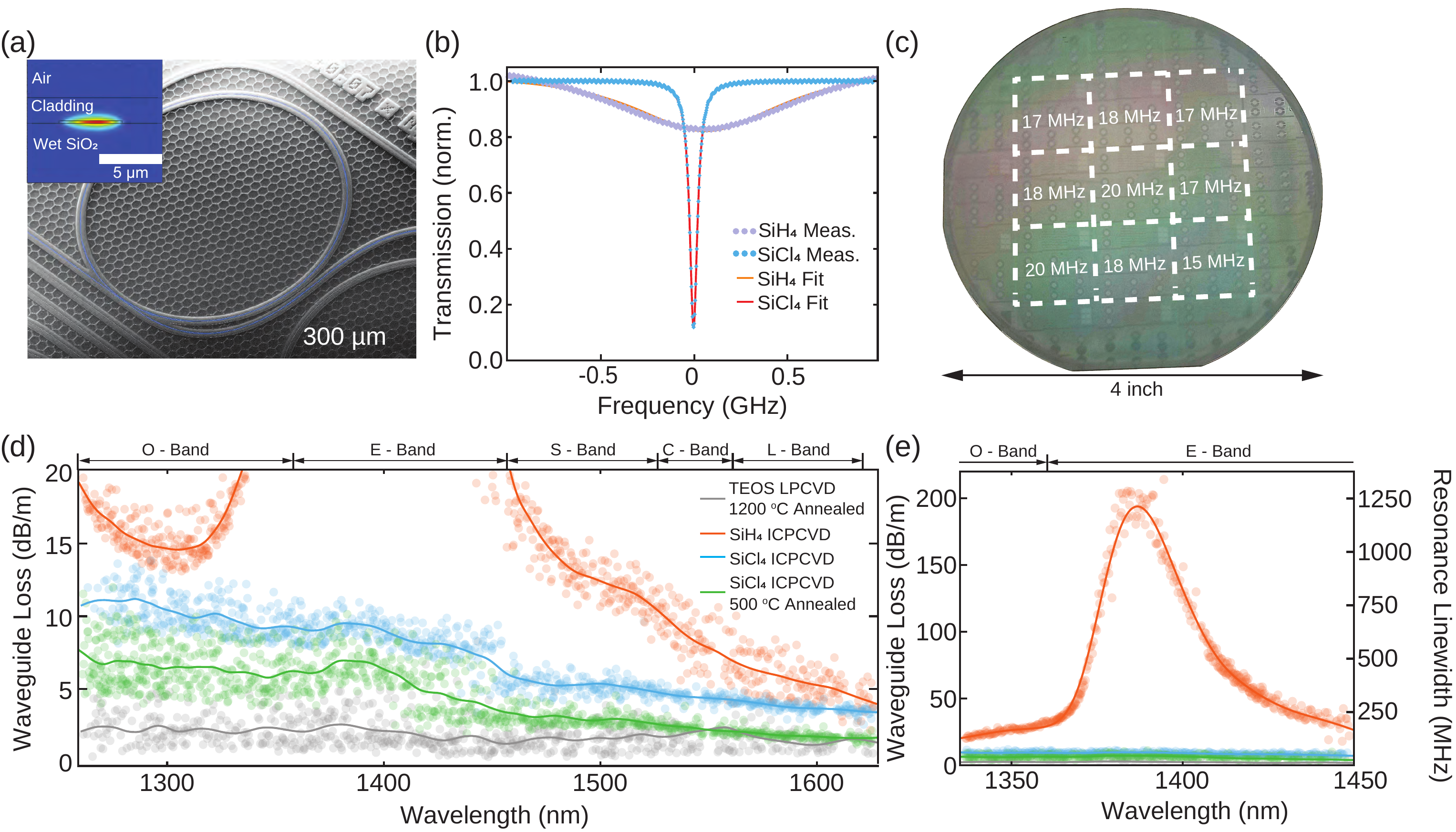}
  \caption{\footnotesize \textbf{Optical loss characterization of the \ce{SiCl4} based \ce{SiO2} cladding.}
  (a) Scanning electron microscopy (SEM) image of the deposited \ce{SiO2} cladding on low confinement \ce{Si3N4} waveguide resonator.
  Inset: simulated mode field distribution in the cladded \ce{Si3N4} resonators.
  (b) Comparison of typical resonance linewidths of resonators cladded by \ce{SiCl4} based \ce{SiO2} and \ce{SiH4} based \ce{SiO2} near the \ce{OH} absorption peak. 
  Center frequencies for the \ce{SiCl4} and the \ce{SiH4} one are 216.88~THz and 216.77~THz, respectively.
  (c) Map of the median intrinsic linewidths around 1550~nm on different stepping fields of a cladded 4-inch wafer (D143\textunderscore 01).
  (d) Waveguide losses of the resonators cladded with annealed LPCVD \ce{SiO2}, \ce{SiH4} based ICPCVD and \ce{SiCl4} based ICPCVD as functions of frequency.
  The optical communication wavelength bands are marked on the axis.
  (e) Waveguide losses of the resonators near the \SI{1380}{\nano\meter} \ce{OH} absorption peak.%
  }
\label{Fig3}
\end{figure*}

\section{Physical characterization of deposited film}

We characterized the infrared absorption (\fref{Fig2}(e)) of the deposited films with an attenuated total reflectance Fourier transform infrared spectrometer (FTIR, Pelkin-Elmer Spectrum 3, diamond UATR) and compared with \ce{SiO2} produced by different methods (see Methods). 
\ce{SiO2} films, deposited with \ce{SiH4} precursor in a basic parallel-plate PECVD tool, show the strong absorption signature of \ce{SiO-H} bond and \ce{Si-H}.
Unannealed LPCVD LTO and TEOS films also feature a wide and strong \ce{SiO-H} absorption peak in $3200$ to $3700~\mathrm{cm^{-1}}$.
The \ce{SiH4} based ICPCVD sample shows a reduced peak absorbance and peak width compared to other samples, while the \ce{SiO-H} absorption peak is still evident.
As expected, the absorbance of \ce{SiCl4} based ICPCVD film is very close to the wet silicon oxide reference without any perceivable hydrogen absorption peak.

X-ray fluorescence spectroscopy indicates the deposited film contains residual \ce{Cl}.
While excessive \ce{Cl} content in the deposited \ce{SiO2} can destabilize the film and make the film hygroscopic, we find that if a high bias RF power is applied during the deposition, the resulting film is not hygroscopic.
The optical loss of cladded resonators can remain stable for more than three months when stored in laboratory conditions.
The remaining \ce{Cl} in \ce{SiO2} also leads to a small increase of refractive index compared to thermal \ce{SiO2} (0.01 at \SI{632.8}{\nano\metre}) (\fref{Fig2}(d)). 
Atomic force microscopy characterization shows a smooth top surface of $<\SI{0.25}{\nano\meter}$ root-mean-square roughness after \SI{1.8}{\micro\meter} deposition.
This smooth surface can lower the scattering loss and is critical for further heterogeneous integration by direct bonding.

\section{Optical loss and applications}

\begin{figure*}[htb]
  \centering
  \includegraphics[width=0.95\textwidth]{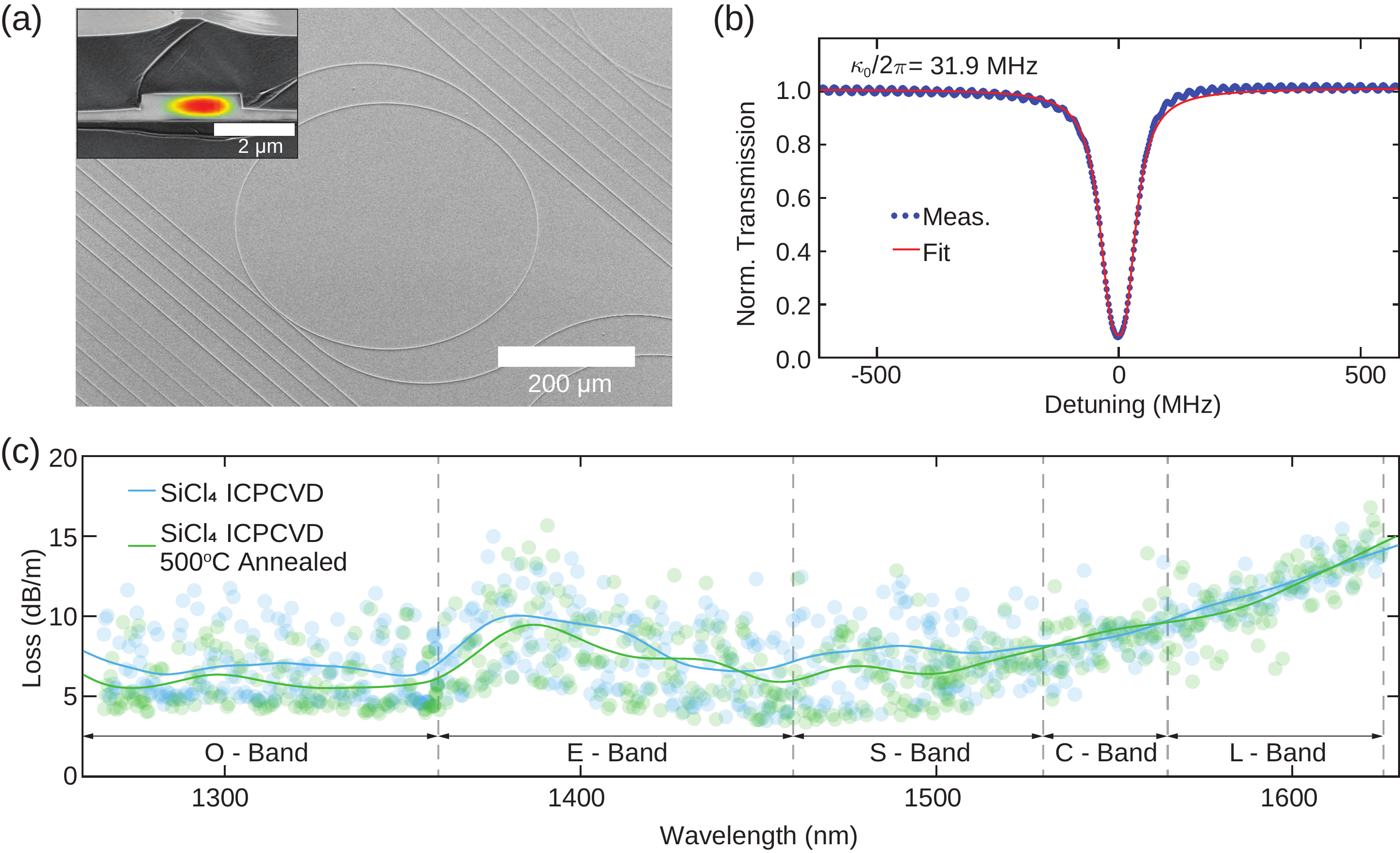}
  \caption{\footnotesize \textbf{Application of \ce{SiCl4} based \ce{SiO2} cladding on thin-film \ce{LiNbO3} devices.}
  (a) SEM image of fabricated \ce{LiNbO3} waveguide ring resonator before cladding deposition. 
  Inset: Cross-section SEM image of cladded \ce{LiNbO3} waveguide ring resonator, overlaid with the simulated mode field.
  (b) A typical resonance of the fabricated resonator at 193.06~THz, with the Lorentzian fitting indicating the 31.9~MHz intrinsic linewidth.
  (c) Wideband waveguide loss characterization of cladded devices, as-deposited (\SI{300}{\degreeCelsius}) and annealed (\SI{500}{\degreeCelsius}).
  }
\label{Fig4}
\end{figure*}
%\hl{Resonance: 193.055 THz, k0/2pi = 31.9 MHz}

To evaluate the material loss of the \ce{SiO2} cladding film, we deposit it on top of \ce{Si3N4} ring resonators as the cladding and compare the measured resonance linewidth with resonators fabricated with established processes\cite{siddharth2023hertz} (see Methods).
The resonance linewidths of the cladded resonator were characterized with a home-built optical vector network analyzer\cite{riemensberger2022photonic} and the waveguide optical loss is computed from the measured intrinsic linewidth (\fref{Fig3}(b)).

We use low confinement \ce{Si3N4} ring resonators with \SI{200}{\nano\metre}$\times$\SI{5}{\micro\metre} waveguide cross-section and \SI{50}{\giga\hertz} free spectral range (FSR).
For the fundamental transverse electrical mode ($\mathrm{TE}_{00}$), $\sim\,23\%$ of total optical intensity is distributed in the top cladding  (Fig.~\ref{Fig3}(a)), making the device ideal for sensing the additional loss caused by the deposited cladding.
We deposited $\sim$\SI{1.85}{\micro\metre} of \ce{SiO2} on top of the waveguides using the \ce{SiCl4} based process.
For comparison, another sample from the same wafer was cladded with \SI{1.77}{\micro\meter} of \ce{SiO2} with \ce{SiH4} based ICPCVD in the same tool (see Methods).
%This reference process is optimized for the reduced \ce{OH} absorption and enhanced gap-filling effect.

As the reference for optical loss evaluation, another \ce{Si3N4} device wafer was cladded by a well-established LPCVD oxide process.
The stack consists of \SI{1}{\um} of oxide deposited from TEOS precursor, and \SI{2}{\um} of oxide deposited with the standard low-temperature oxide (LTO) process using \ce{SiH4} precursor.
This reference wafer was annealed at \SI{1200}{\degreeCelsius} for \SI{11}{\hour} after both TEOS and LTO deposition to fully remove hydrogen impurities and densify the film.
Although the baseline LPCVD cladding process is known to yield a very low material loss, the high temperature of the TEOS-based deposition and the extensive annealing seriously limits its compatibility with advanced materials and other processing steps.

\Fref{Fig3}(d), (e) shows the loss spectrum of the cladded waveguides.
While the one with ICPCVD film deposited with \ce{SiH4} and \ce{O2} shows a $200~\mathrm{dB/m}$ absorption peak at \SI{1380}{\nm}, the hydrogen absorption peak is indiscernible for the waveguide using the \ce{SiCl4} based film.

Conservatively including the higher scattering loss of the thinner ICPCVD film, the estimated material loss of the as-deposited \ce{SiCl4} based \ce{SiO2} film is $<$\SI{9}{\dB\per\metre} higher than the high-temperature LPCVD process at \SI{1550}{\nm}.
A low waveguide loss is measured across the entire characterization range of \SI{1260}{\nano\meter} to \SI{1625}{\nano\meter}, limited by the availability of tunable lasers.
The higher loss at shorter wavelengths may be attributed to scattering loss at the top surface of cladded waveguides or defects in the deposited film.

We also note that the \SI{1550}{\nm} to \SI{1620}{\nm} loss of cladded \ce{Si3N4} waveguides can be further reduced to $< \SI{2.5}{\dB\per\metre}$ after an \SI{1}{\hour} short annealing at \SI{500}{\degreeCelsius}, which leads to no measurable additional optical loss to the ones with high-temperature LPCVD cladding.
This loss reduction may be partially caused by the healing of \ce{Si3N4} damage due to the intense ultra-violet light generated by the plasma discharge in the deposition\cite{UV}.

To demonstrate the compatibility of the novel ICPCVD process with \ce{LiNbO3} on insulator (LNOI) platform, we cladded a thin-film \ce{LiNbO3} photonic integrated circuits wafer with our \ce{SiO2} film and evaluated the optical loss.
A commercial Z-cut LNOI wafer (NanoLN) with \SI{600}{\nano\meter} thick \ce{LiNbO3}, \SI{4.7}{\micro\meter} thick buried \ce{SiO2} is etched with a diamond-like-carbon hardmask\cite{li2023high}.
We etched \SI{400}{\nano\meter} of \ce{LiNbO3} for the ridge waveguide and leave a \SI{200}{\nano\meter} slab (\fref{Fig4}(a)).
%We then
Then we deposited an \SI{1.7}{\micro\meter} \ce{SiCl4} based \ce{SiO2} film and characterized the resonance linewidths of \SI{80}{\giga\hertz} FSR racetrack resonators to derive the waveguide loss (\fref{Fig4}(b)).
The loss of the cladded waveguides is comparable with uncladded devices in the earlier report\cite{li2023high}.
The sample was then annealed in oxygen at \SI{500}{\degreeCelsius} for \SI{1}{\hour} and the measured optical loss shows a slight decrease (\fref{Fig4}(c)).

\section{Summary}

In summary, our work presents a novel ICPCVD process, allowing low-temperature deposition of optical grade \ce{SiO2} cladding film, with very low near-infrared loss in the range of \SI{1260}{\nano\meter} to \SI{1625}{\nano\meter}, which covers the entire telecom S, C, and O band without a gap. 
Our process is demonstrated on a commercially available tool and is fully foundry compatible.
As an essential building block for the fabrication process, we believe this work can unlock the full potential of today's low-loss integrated photonics platforms with low thermal budgets such as \ce{LiNbO3} on insulator, erbium-doped \ce{Si3N4} and III-V group semiconductor photonic integrated circuits.

\section*{Methods}
\begin{footnotesize}

\noindent\textbf{Fabrication of the \ce{Si3N4} ring resonators}:

The resonators were fabricated by depositing \SI{200}{\nano\meter} stoichiometric \ce{Si3N4} on silicon wafers with $>$\SI{7.2}{\um} wet oxide layer by LPCVD.
\ce{Si3N4} waveguides were directly defined by deep ultra-violet lithography (ASML PAS5500/350C, JSR M108Y, Brewer DUV42P) and flurorine chemistry reactive ion etching.
The devices were annealed in \ce{N2} at \SI{1200}{\degreeCelsius} for \SI{11}{\hour} before cladding deposition.

\noindent\textbf{Deposition condition of the \ce{SiO2} films}: 

\ce{SiCl4} based ICPCVD (Run 108): Oxford Instruments PlasmaPro 100, 50~sccm \ce{O2}, 10~sccm \ce{Ar}, 30~sccm \ce{SiCl4}, 5~mTorr, 2000~W ICP RF power (2~MHz), 250~W bias power (13.56~MHz), 1980~s. 
Our experience shows, that in order to achieve high quality film with low optical loss and low etch rate in hydrofluoric acid, a high ICP power above \SI{400}{\watt} is needed.

\ce{SiH4} based ICPCVD (Run 127): Oxford Instruments PlasmaPro 100 (same machine as used for \ce{SiCl4} based samples), Deposition step 1: 33~sccm \ce{O2}, 30~sccm \ce{Ar}, 24~sccm \ce{SiH4}, 2~mTorr, 2000~W ICP RF power, 250~W bias power, 420~s; Deposition step 2 : 17~sccm \ce{O2}, 15~sccm \ce{Ar}, 12~sccm \ce{SiH4}, 1~mTorr, 2000~W ICP RF power, 350~W bias power, 900~s; Deposition step 3: same as step 2 but with 250~W bias power, 1410~s.

\ce{SiH4} based PECVD: Oxford Instruments PlasmaLab 100, \SI{300}{\degreeCelsius}, 1~Torr, 400~sccm 2\%\ce{SiH4}/\ce{Ar}, 710~sccm \ce{N2O}, 20~W capacitively coupled RF (13.56~MHz) power.

\noindent\textbf{Mitigation of residual hydrogenous gas contamination}:

We note that residual gas in the process chamber, coming from the previous hydrogenous deposition or the hydrous air when the wafer is loaded to the reactor, can lead to hydrogen contamination in the deposited film and elevated \ce{OH} absorption.
As the system is designed for the requirements in semiconductor processing, the deposition chamber has no provisions for high vacuum and can hardly be pumped to $<10^{-7}~\mathrm{Torr}$.
It's inevitable that traces of hydrogen and water -- the most common residual gas in vacuum systems -- will remain in the chamber.
Thus, careful chamber conditioning by dummy processes and complete dehydration of substrates before depositions are necessary to minimize the amount of residual hydrogen.
Mass spectroscopy residual gas analyzers are the ultimate instruments for the evaluation of residual gas in high vacuum systems, however, it's not compatible with our deposition tool due to technical limitations.
An economical alternative is doing a in-situ atomic emission spectroscopy by starting a \ce{Ar} plasma in the process chamber and monitoring optical emission at \SI{656}{\nano\meter} from excited \ce{H} atoms with a optical spectrometer. (See Supplementary)
Further conditioning deposition and degassing may be necessary if the \ce{H} is detected with the spectrometer.
Higher \ce{SiCl4} to \ce{O2} gas flow ratio is also empirically found to be beneficial for reduced residual hydrogen contamination.

\noindent \textbf{Funding Information}:  
This work was supported by the EU H2020 research and innovation program under grant No.965124 (FEMTOCHIP) and by the Horizon Europe EIC Transition program under grant no. 101113260 (HDLN).
It was further supported by the Air Force Office of Scientific Research (AFOSR) under Award No. FA9550-19-1-0250.

\noindent \textbf{Acknowledgments}: 
The samples were fabricated in the EPFL center of MicroNanoTechnology (CMi). 
We thank technician Cuenod Blaise for assistance in the process development %
and Michael Zervas for the early stage work for the acquisition of the ICPCVD tool.
We thank Guilherme Almeida for insightful discussions on the chemical thermodynamics.

\noindent \textbf{Author contributions}: 
Z.Q. and Z.L. performed the deposition experiments and process development with the assistance of R.N.W..
A.S. and Z.L. designed the test resonators used in experiments.
R.N.W., Z.Q., X.J. fabricated the \ce{Si3N4} devices used in experiments.
R.N.W. fabricated the silicon test structures for gap-filling experiments.
Z.L. fabricated the LNOI devices.
Z.L. and Z.Q. carried out data analysis and simulations. 
M.D. assisted in the procurement and installation of the ICPCVD tool, and aided the design of experiments.
Z.Q. wrote the manuscript with the assistance from Z.L. and input from all co-authors.
T.J.K supervised the project.

\noindent \textbf{Data Availability Statement}: 
The code and data used to produce the plots within this work will be released on the repository \texttt{Zenodo} upon publication of this preprint.

\noindent \textbf{Competing interests}
T.J.K. is a cofounder and shareholder of LiGenTec SA, a start-up company offering \ce{Si3N4} photonic integrated circuits as a foundry service as well as Luxtelligence SA, a foundry commercializing \ce{LiNbO3} photonic integrated circuits.

\end{footnotesize}
%\vspace{-0.3cm}

%% NOTE: RevTeX only compiles with BibTeX, Configure your system to use it ONLY!

\renewcommand{\bibpreamble}{%
$\dagger$These authors contributed equally to this work.\\
$\ast$\href{mailto://tobias.kippenberg@epfl.ch}{tobias.kippenberg@epfl.ch}
}
\pretolerance=0
\bigskip
\bibliographystyle{apsrev4-2}
\bibliography{library}
\end{document}

\title{Supplementary information for: \\ Hydrogen-free low temperature silica for next generation integrated photonics}
%Low-temperature deposition of low-loss silica cladding\\ for next-generation integrated photonics}

\author{Zheru Qiu$^{1,2,\dagger}$, Zihan Li$^{1,2,\dagger}$, Rui Ning Wang$^{1,2,3}$, %
Xinru Ji$^{1,2}$, Marta Divall$^{1,2}$, Anat Siddharth$^{1,2}$, Tobias J. Kippenberg$^{1,2,\ast}$}

\address{${}^1$Swiss Federal Institute of Technology Lausanne (EPFL), CH-1015 Lausanne, Switzerland\\
${}^2$Center for Quantum Science and Engineering, EPFL, CH-1015 Lausanne, Switzerland\\
${}^3$Present Address: Luxtelligence SA, CH-1015, Lausanne, Switzerland}

%\begin{abstract}
%\textbf{.}
%\end{abstract}

%%%%%% RESET EQUATION NUMBERS ETC. %%%%%%%%
\setcounter{equation}{0}
\setcounter{figure}{0}
\setcounter{table}{0}

\setcounter{subsection}{0}
\setcounter{section}{0}

\maketitle
\vspace{-4em}
{\hypersetup{linkcolor=black}\tableofcontents}
%\newpage

%%%%%%%%%%%%%%%%%%%%%%%%%%%%%%%%%%%%%%%%%%%%%%%%%%%%%%%%%%%%%%%

\section{Details on the thermodynamics analysis of the oxidation reactions}

%We compared the  is a result of the significantly higher bond energy of Si-Cl than Si-H and the lower stability of \ce{Cl2} than \ce{H2O}.

The full description of the thermodynamic tendency of a chemical reaction is given by the reaction Gibbs free energy change $\Delta G=\Delta H-T\Delta S$ at the specified temperature.
We are not running the depositions at the standard condition but at higher temperatures and lower pressures, here we provide the justifications for using the standard reaction enthalpy change to describe the tendency for convenience.
Noting that neither of the oxidation of \ce{SiCl4} and \ce{SiH4} change the number of molecules in the gas phase, the $\Delta G$ is dominated by the reaction enthalpy $\Delta H$ change, as the $T\Delta S$ term is significantly smaller than $\Delta H$.
This is further confirmed by the standard thermodynamic properties table in reference \cite{CRC}, which gives $\Delta S^\circ = \SI{9.3}{J/mol/K}$ for each mole of \ce{SiH4} and $\Delta S^\circ = \SI{-48.2}{J/mol/K}$ for each mole of \ce{SiCl4}.
These $\Delta S$ terms contribute to less than \SI{35}{kJ/mol} in the temperatures of 273 K to 573 K, insignificant in comparison with the \SI{1136}{kJ/mol} difference in $\Delta H$.
Concerning the relation of $\Delta H$ with temperature, as $\partial H/\partial T = C_p$, where $C_p$ is the heat capacity under constant pressure.
The contribution of $C_p$ is also negligible according to the data in reference \cite{CRC}.
%Delta Cp=-7.5 J/mol/K for SiH4
%−39.4 J/mol/K for SiCl4

\section{Details on the optical emission spectroscopy for residue hydrogen detection}

\begin{figure}[htbp]
	\centering
	\includegraphics[width=0.9\textwidth]{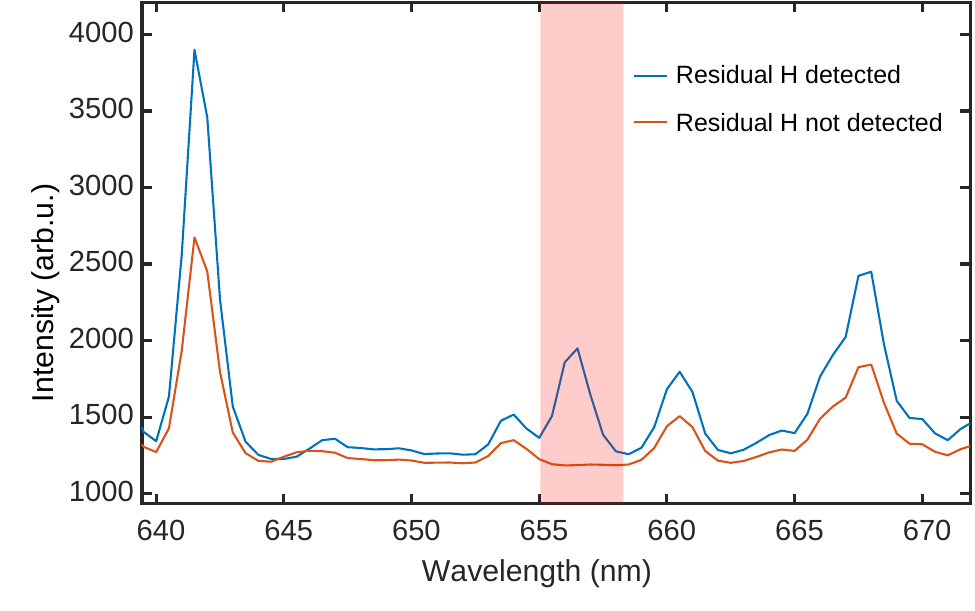}
	\caption{\textbf{Optical emission spectrum}.
		Measured optical emission spectrum near \SI{656}{\nano\meter} hydrogen emission line.
		}
	\label{Fig:OES}
\end{figure}

To characterize the residual hydrogenous gases in the chamber, we started a \ce{Ar} plasma discharge at the condition of 3~mTorr, 2000~W ICP power, and 30~sccm of \ce{Ar} gas flow.
The chamber was monitored through the side window (after the RF shielding mesh and UV filter) by a fiber-coupled high-throughput optical spectrometer.
The emission spectrum is shown in \Fref{Fig:OES} for two typical conditions where residual hydrogen is detected and not detected.

\section{X-ray fluorescence spectroscopy analysis of deposited film}

\begin{figure}[htbp]
	\centering
	\includegraphics[width=0.9\textwidth]{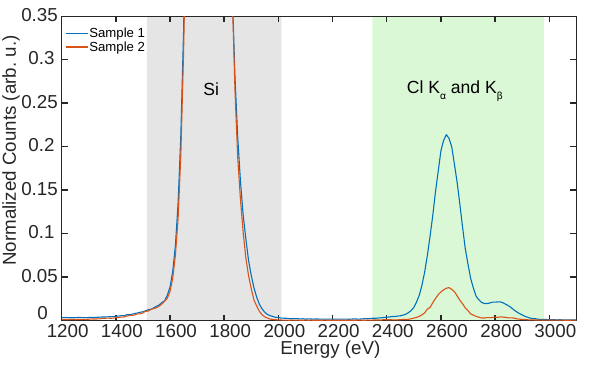}
	\caption{\textbf{X-Ray fluorescence spectrum}.
        Spectrum of the two \ce{SiO2} samples deposited in two different conditions on \ce{Si} substrate.
		}
	\label{Fig:XRF}
\end{figure}

We characterized the deposited \ce{SiO2} film with X-ray fluorescence Spectroscopy (XRF).
Although we cannot quantify the elemental composition of the film due to the lack of a standard sample, we compared the chlorine content of the film deposited under different conditions.
The X-ray count rate was normalized by the film thickness measured by optical reflectometry and the count rate of \ce{Si} from the substrate.
Sample 1 was deposited under the condition of 10~mTorr, 30~sccm \ce{SiCl4}, 50~sccm \ce{O2}, 0~sccm \ce{Ar}, 100~W RF bias, 2000~W ICP power, and \SI{300}{\degreeCelsius} (Run 95).
Sample 2 was deposited under the condition of 6~mTorr, 30~sccm \ce{SiCl4}, 50~sccm \ce{O2}, 10~sccm \ce{Ar}, 350~W RF bias, 2000~W ICP power, and \SI{300}{\degreeCelsius} (Run 103).
As shown in \Fref{Fig:XRF}, the chlorine content of sample 1 is 5.7 times higher than sample 2, which may be attributed to the higher \ce{SiCl4} concentration in the reactor.
Sample 1 is not stable in long-term storage while sample 2 is stable.

\clearpage
%%%%%%%%%%%%%%%%%%%%%%%%%%%%%%%%%%%%%%%%%%%%%%%%%%%%%%%%%%%%%%%%

%\section*{Supplementary References}
%\bigskip
\renewcommand{\bibpreamble}{
$\dag$These authors contributed equally to this work.\\
$\ast$\textcolor{magenta}{tobias.kippenberg@epfl.ch}
}
\bibliographystyle{apsrev4-2}
\bibliography{library}